# Interactive exploration of population scale pharmacoepidemiology datasets

Tengel Ekrem Skar[1], Einar Holsbø[1], Kristian Svendsen[2], Lars Ailo Bongo[1*]

[1] Department of Computer Science, UiT The Arctic University of Norway
[2] Department of Pharmacy, UiT The Arctic University of Norway

* Corresponding author: lars.ailo.bongo@uit.no

## Abstract

Population-scale drug prescription data linked with adverse drug reaction (ADR) data supports the fitting of models large enough to detect drug use and ADR patterns that are not detectable using traditional methods on smaller datasets. However, detecting ADR patterns in large datasets requires tools for scalable data processing, machine learning for data analysis, and interactive visualization. To our knowledge no existing pharmacoepidemiology tool supports all three requirements. We have therefore created a tool for interactive exploration of patterns in prescription datasets with millions of samples. We use Spark to preprocess the data for machine learning and for analyses using SQL queries. We have implemented models in Keras and the scikit-learn framework. The model results are visualized and interpreted using live Python coding in Jupyter. We apply our tool to explore a 384 million prescription data set from the Norwegian Prescription Database combined with a 62 million prescriptions for elders that were hospitalized. We preprocess the data in two minutes, train models in seconds, and plot the results in milliseconds. Our results show the power of combining computational power, short computation times, and ease of use for analysis of population scale pharmacoepidemiology datasets. The code is open source and available at: https://github.com/uit-hdl/norpd_prescription_analyses

## Introduction

Pharmacoepidemiology is the study of the use of and the effects of drugs and other medical devices in large numbers of people [1]. An important effect studied in pharmacoepidemiologic studies is the risks of adverse effects such as hospital admissions coming from using drugs [2]. Since there are many possible drug combinations such studies require very large cohorts to find any signals of adverse drug reactions (ADR), but also to adjust these findings with drug consumption rates. Our power calculations show that to study the difference in reporting rates for an ADR reported in 1 in 50000 drug users (normal reporting rate) between two drugs used by 10% and 2% of the population respectively, there is a need for data covering around 70 million people [3].

Consumption-adjusted ADR rates have only been estimated in smaller studies of relatively few individual drugs [4], [5]. We aim to provide such rates based on population-scale datasets with billions of prescriptions. We will use a retrospective cohort study design that combines prescription register data with spontaneous reports of drug combinations with suspected ADR. We aim to collect this data from several European countries covering a significant part of the population of Europe. Our aim is to use these largely unexploited datasets to explore how drugs are consumed by different populations in the data, and develop analyses of drug interactions and correlations. This is an exploratory, hypothesis-generating study.

We observe three key requirements for exploring such large datasets. First, to enable interactive exploration the analysis time must be short, preferably less than 5 seconds. These big data computations therefore require large amounts of computing power, and thus the use of distributed computing, and hardware acceleration such as GPUs. Second, discovering patterns in the data requires developing and evaluating novel machine learning and statistical methods. It is therefore necessary to use machine learning frameworks. Third, the patterns must be interpreted. We therefore need interactive visualization tools. Summarized, we need an approach that provides intuitive ways to transform and interactively analyze the big datasets by abstracting away unnecessary details.

To our knowledge, no existing pharmacoepidemiology tool satisfies all three requirements. Most studies use the classic epidemiological approach where they do statistical tests for a hypothesis about outcomes for specific exposures. We use an agnostic approach without any prior hypothesis, to find novel drug interactions and associated outcomes. We must therefore process a large dataset such that it can be explored with short response times. We use state-of-the-art big data processing and machine learning frameworks to implement novel data exploration methods. Big data processing frameworks such as Spark [6] are designed to utilize the aggregated computation power, RAM size, and I/O bandwidth of distributed computing to process low-dimensional data such as our ADR and prescription datasets. We can therefore utilize the computational power of distributed computers and GPUs to reduce the computation time for the analyses to a few seconds, and to scale to very large datasets. Machine learning frameworks such as Keras (https://keras.io/) or Tensorflow [7] provide the computation power needed for deep learning methods, whereas scikit-learn [8] provides large libraries of machine learning and statistical methods. Finally, interactive scripting environments such as Jupyter Notebook (https://jupyter.org/) and RStudio (https://rstudio.com/) enable swift implementation of scripts and visualizations, which makes these excellent for data exploration and interpretation.

We have developed a data transformation pipeline, data structures, machine learning methods and visualizations for exploring drug usage patterns and associated outcomes. These can be adapted for other analyses using an interactive programming environment that enables quick prototyping and iterative improvements. We have used our system to integrate and explore three datasets. Two are from the Norwegian Prescription Database: 370 million prescriptions covering the entire Norwegian population and contains most prescriptions at pharmacies from 2004-2014, and 60 million prescriptions with additional variables that cover Norwegian elders (>65 years). The third is a dataset with 1.9 million hospitalization records from the Norwegian Patient Registry.

# Design and implementation

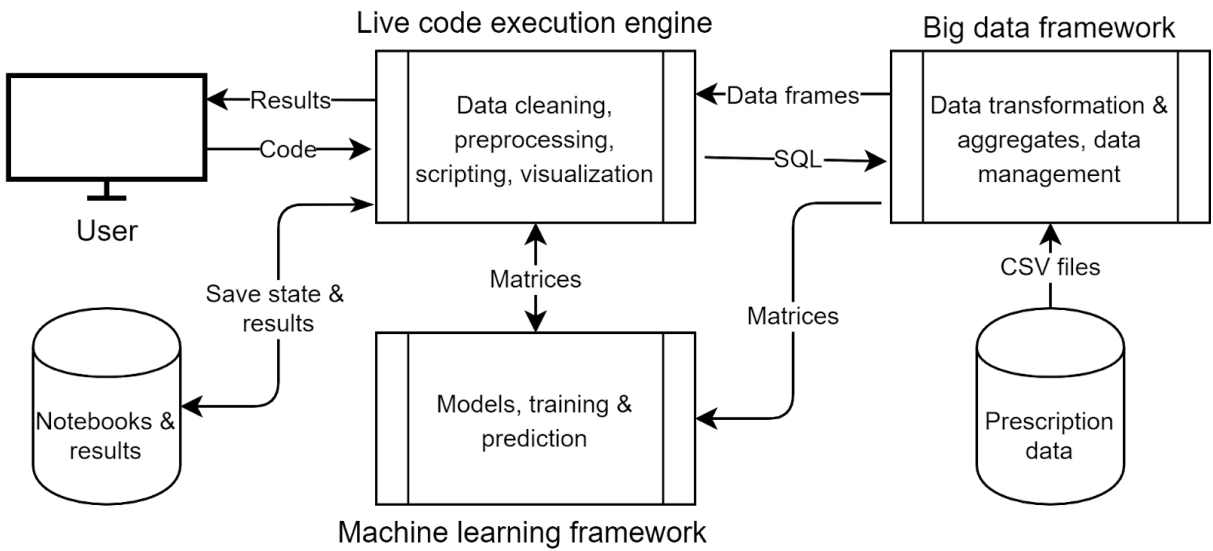

*Figure 1: System architecture.*

The system architecture is centered around the Jupyter live code execution engine where the user writes code in a programming environment and interacts with the resulting visualizations (Figure 1). The engine connects all frameworks, tools and libraries. It enables the user to leverage multiple frameworks and libraries, and therefore implement and execute the entire workflow in a single environment. In addition to reducing the development time, analysis time can also be reduced since code changes do not require re-executing the entire analysis.

We store the prescription data in the Spark big data framework. The user transforms, aggregates, and queries the data using SQL queries. The transformed data can be further cleaned and preprocessed using library functions executed in the live code execution engine. The user programs methods to train, tune, and validate machine learning models for the preprocessed data. These are executed in the Keras or scikit-learn machine learning frameworks. Finally, the user interprets the results using visualization and other libraries in the live execution engine. The results are stored as notebooks with all code and visualizations.

## Data cleaning and transformation using Spark

We use the Apache Spark [6] (https://spark.apache.org/) with Spark SQL's DataFrame API for data cleaning and transformation. Spark has quickly become the industry standard for big data processing and it is offered by most cloud platforms, and it supports interactive queries on large datasets [6]. Our prescription dataset with hundreds of millions of rows and few columns is well suited for Spark. We therefore use Spark SQL to implement data transformation and aggregation. It has four main advantages. First, queries are only a few lines of code. For example, transformations can be expressed with group-by on one or multiple columns, and then applying an aggregate function to the grouped data. Second, the schemas can be used interactively. For example, we can check the state of the data between transformations by printing the schema or displaying a subset of the data. Third, we do not need to keep track of the data type of each element with every transformation, nor the state of the data. Fourth, Spark SQL provides a query optimizer.

## Drug use pattern recognition using scikit-learn

We use scikit-learn (https://scikit-learn.org/) and Keras CIT (https://keras.io/) on Tensorflow [7] to implement, train, and evaluate our drug use pattern recognition algorithms. Sci-kit provides a large number of algorithms for preprocessing, and supervised and unsupervised learning. Keras is a high-level API for building and training deep learning models, which provides a simplified programming interface for designing deep learning models. Tensorflow supports GPU acceleration and distributed training.

## Explorative analyses using Jupyter

We use Jupyter Notebook (https://jupyter.org/) for data exploration and interpretation. It supports many programming languages, including Python, Scala, Java, Go. Jupyter is increasingly used and supported by the data science community. The user interacts with a web app connected to a Jupyter web server that executes the code in a kernel. The user writes blocks of code in cells, but the kernel maintains the notebook's state between each cell. The results for an executed cell are displayed in an output cell below the code block. The cells also support other formats, such as raw text or MarkDown (https://daringfireball.net/projects/markdown/), so notes and methodology descriptions can coexist with the code. The resulting code, visualizations, and documentation is the notebook can be saved as a file that can be executed to reproduce the analysis.

We use the Spylon Jupyter kernel (https://github.com/Valassis-Digital-Media/spylon-kernel) to enable seamless transition between writing efficient preprocessing and computations in Scala, and interpretation in Python. It initializes a Spark Session on startup. The Spark Session connects to an existing Spark cluster if Spark is configured, otherwise it creates a virtual Spark cluster that runs on the same machine as the Jupyter environment.

We use the Spark DataFrame API, which provides support for SQL querying and registration of temporary tables. The Spark Session is shared between the interpreters, so the tables are also accessible from Python. We also extract and convert Spark Dataframe into Panda Dataframes (https://pandas.pydata.org/), so we can use these in Python for efficient queries on transformed and aggregated data (an example is in Figure 2). For visualization we use matplotlib (https://matplotlib.org/) and Plotly. We use the Python 3 kernel since Spylon has limited support for Plotly (https://github.com/plotly/plotly.py).

```python
# Get count of drugs that were prescribed to men who are alive
men_a_dc = spark.sql("select drugcode, count as count_a from
men_a_drug_c").sort("drugcode").where("count>50").toPandas()#.set_index(
"drugcode")

# Get count of drugs that were prescribed to men that (later) died
men_d_dc = spark.sql("select drugcode, count as count_d from
men_d_drug_c").sort("drugcode").where("count>23").toPandas()#.set_index(
"drugcode")

# Merge the two tables (living and dead), indexed by drug code
# with two additional columns (drugcount of patients who are alive,
# and of patients who are dead).
# Drop rows (drugs) which are not prescribed in the live and dead population
combined = men_a_dc.merge(men_d_dc,
on="drugcode",how="inner").sort_values("drugcode").fillna(0.0).reset_index(
drop=True)
```

Figure 2: Example from Supplementary Notebook 2. Spark-SQL is used to extract drug counts for dead and alive patients. The results are converted to Panda frames that are inner joined such that only the drugs present in both datasets are included.

# Case study: exploring drug use patterns and hospitalization

An important goal in pharmacoepidemiology is finding and understanding adverse drug reactions (ADR) for combinations of drugs. For such studies the elderly population is especially interesting, since they have the highest drug consumption, and they are comedicated with many drugs at the same time. The elderly therefore have most ADR and the resulting health issues, unnecessary hospitalization, and death [9]. The challenge is detecting the cause for ADR for rarely prescribed comedication. In this case study, we demonstrate how large prescription data can be integrated and analyzed in a modern workflow. We explore drug consumption patterns that lead to hospitalization by finding drugs that are over-represented in elderly hospitalized patients. This is the first step towards finding ADRs for comedicated drugs, generating hypotheses for further analyses.

## Drug usage datasets and variables

*Table 1: The three datasets used in this case study.*

| Name | Samples | # drugs | details |
| --- | --- | --- | --- |
| General population | 374.9M | 856 | Prescriptions in the period 2004-2014, NorPD. Representative of drug consumption in the general population. |
| Elders | 61.9M | 1200 | Prescriptions 2012-2014, NorPD. Patients > 65 years of age. |
| Patient registry | 1.9M | NA | Hospitalization and diagnosis data for the people in the Elders data set. |

Drugs are encoded using The Anatomical Therapeutic Chemical Classification System (https://www.whocc.no/atc_ddd_index/). The ATC codes are organized in a hierarchy with five levels that describe for each drug the organ and disease the drug treats, as well as the pharmacological target and chemical grouping in increasing details down to the specific chemical compound. Drugs that are chemically similar or used similarly share up to four levels in the ATC hierarchy. We use the ATC code hierarchy to adjust the dimensionality of the analyses.

Disease and other health outcomes are coded with ICPC-2 (https://ehelse.no/kodeverk/icpc-2e--english-version) and ICD-10 (https://icd.who.int/en) codes. The International Classification of Primary Care (ICPC-2) is commonly used by general practitioners in Norway. The International Statistical Classification of Diseases and Related Health Problems (ICD) is proposed by the World Health Organization as a global classification system. In Norway, ICD-10 is widely adopted in hospitals.

## Ingest and transform raw data to enable SQL queries

To initialize the data structures we run data ingestion programs to convert the .csv data to more suitable formats (Supplementary Notebook 1). First, we load the csv formatted data into Spark using Spark Session. This automatically parses the field names, and generates a schema for the data. We convert timestamps to unix-timestamp format and convert the birth year to integer format, then save the dataset back to disk in the Parquet format. Spark SQL automatically converts the data to a column format, builds various indexes, and partitions the data to multiple files. We cache these Spark SQL tables, which stores the data and results in RAM to reduce query time.

## Data exploration using SQL queries

We start by comparing the drugs prescribed to patients that died during the 3 year period with those that did not. We split the elders data into two subsets: patients that died between 2013 and 2017 (133.000) and those that were alive (640.000). In each group we count the number of times each drug is prescribed and the total number of prescriptions (7.5 million for dead patients, 16.2 million for alive patients). We normalize the aggregated prescriptions by dividing each prescription count by the total number of prescriptions in each respective group. The resulting table contains the relative frequency of each drug per split.

We compare the relative frequencies from the distribution of drug consumption of the two groups on the log scale. We choose the five drugs with the highest frequency, and then visualize the prescriptions of these drugs in dead patients as an aggregate in a 900 day time frame prior to each patient's death. We aggregate consumption in 30-day windows by each patient's approximate days until death when the prescription occurs.

The results (Figure 3) are as should be expected. Midazolam (N05CD08) is almost exclusively prescribed a very short period prior to death. It is a sedative, and is most likely used for palliation. L01AX03, L02BB04, L01CA04 are all drugs used in cancer treatment, and we see that the prescriptions tend to occur more frequently when closer to patients death. Finally, dexamethasone (H02AB02) is used together with cancer treating drugs, specifically to treat or reduce the severity of adverse effects from the cancer medication, which tend to appear when the treatment is prolonged.

## Data preprocessing to binarize, clean and reduce data dimensionality

To prepare the *elders* data for modeling we encode drug consumption as a binary variable, True or False. First, True for all ATC codes used in the period of interest. Second, True if the person was hospitalized in the period. Otherwise, False. We then split the data by sex since some drugs are sex-specific. We remove patients with chronic disease, since they are often hospitalized because of their disease rather than drug usage. We consider as chronic those patients who had more than 10 hospitalizations between the years 2012 and 2014.

For some analyses we adjust dimensionality using the hierarchical ATC codes. The lowest level (5) has >900 codes, while level 4 only has 160 codes. By using level 4 instead of 5 we lose information about specific chemical compounds prescribed, but retain the chemical group to which the drug belongs and the function of its prescription. The resulting vectors are sparse, since on average only 4 of the level 4 codes are non-zero.

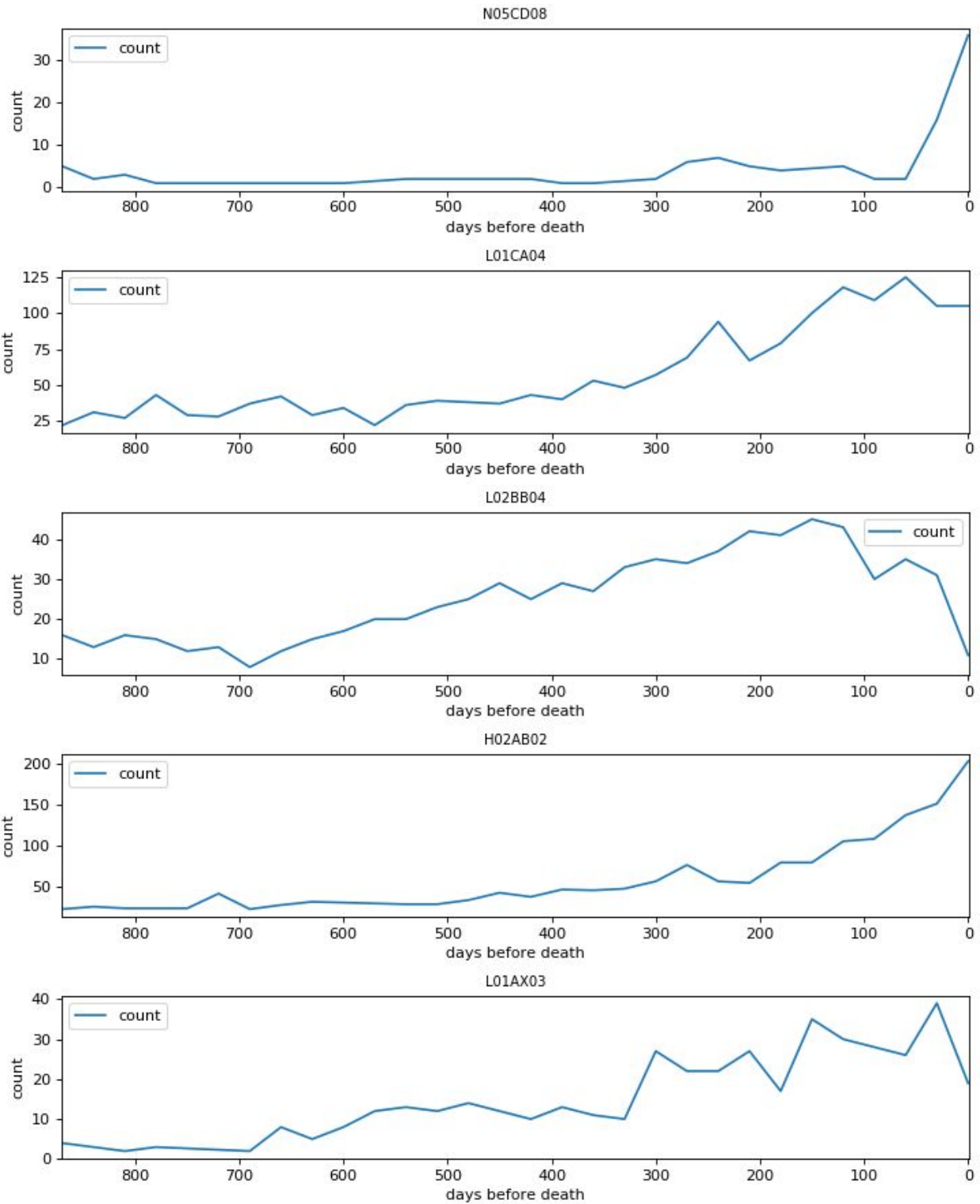

*Figure 3*: Norwegian drug prescription occurrence (30 day aggregates). The code for generating the figure is in Supplementary Notebook 3.

# Differential drug use in hospitalized patients

We apply a logistic regression model to investigate whether drugs are differentially used prior to hospitalization. Here we use 5-level ATC codes. We train the models using drug consumption estimated for 30 day windows for hospitalized and non-hospitalized patients.

For hospitalized patients, we use the 30 day windows immediately before each hospitalization. To reduce the likelihood that our sampled dates are correlated strongly with any previous hospitalization, we only use windows where the period starts at least 60 days after the last hospitalization. For non-hospitalized patients, we randomly select 30 day windows in 2013. We consider a drug as used if it occurs somewhere in the 30-day window (Figure 4).

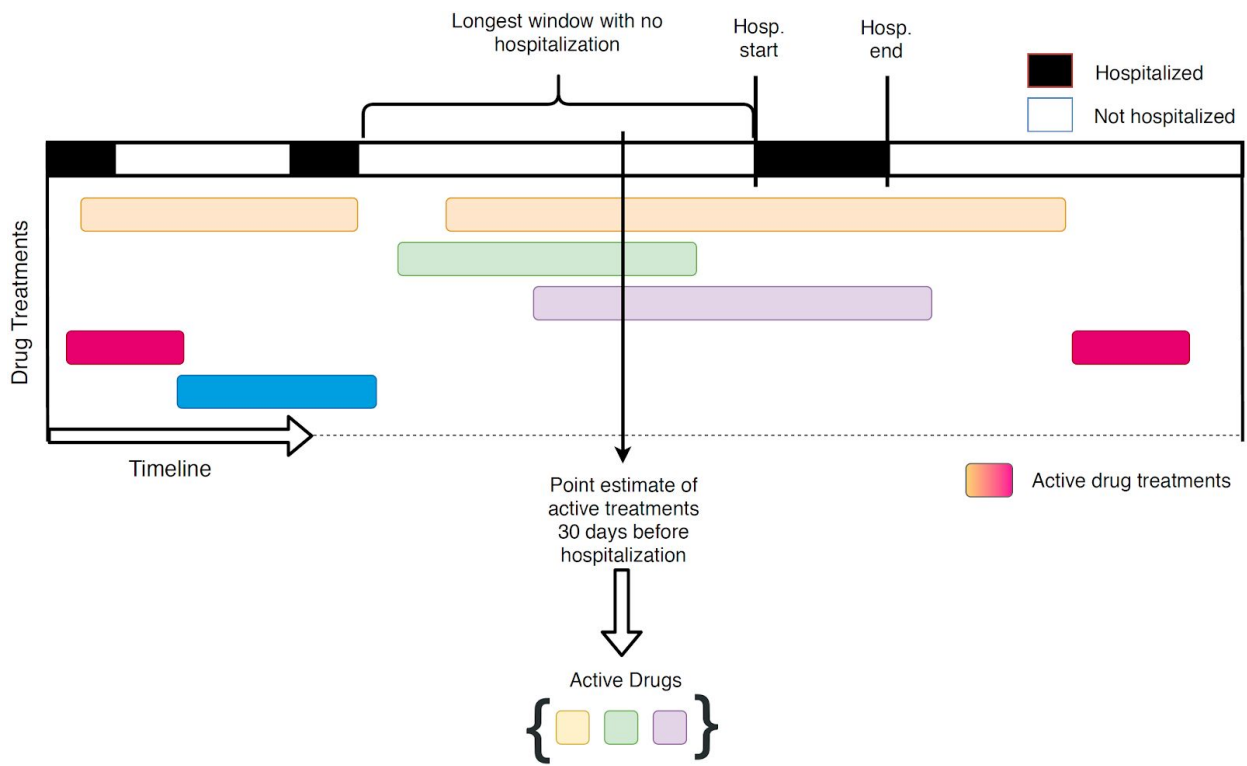

*Figure 4: Drug consumption estimation.*

We provide uncertainty intervals with the bootstrap [10], drawing 2000 bootstrap samples. Figure 5 shows estimated log odds ratios for hospitalization in terms of each active drug. I.e., the odds associated with hospitalization for someone using the drug in the upper left corner is roughly ten times higher than that of someone not using the drug. Hence these estimates can be used for a list of likely suspects. Error bands are quite wide, suggesting that more data is needed for accurate parameter estimates.

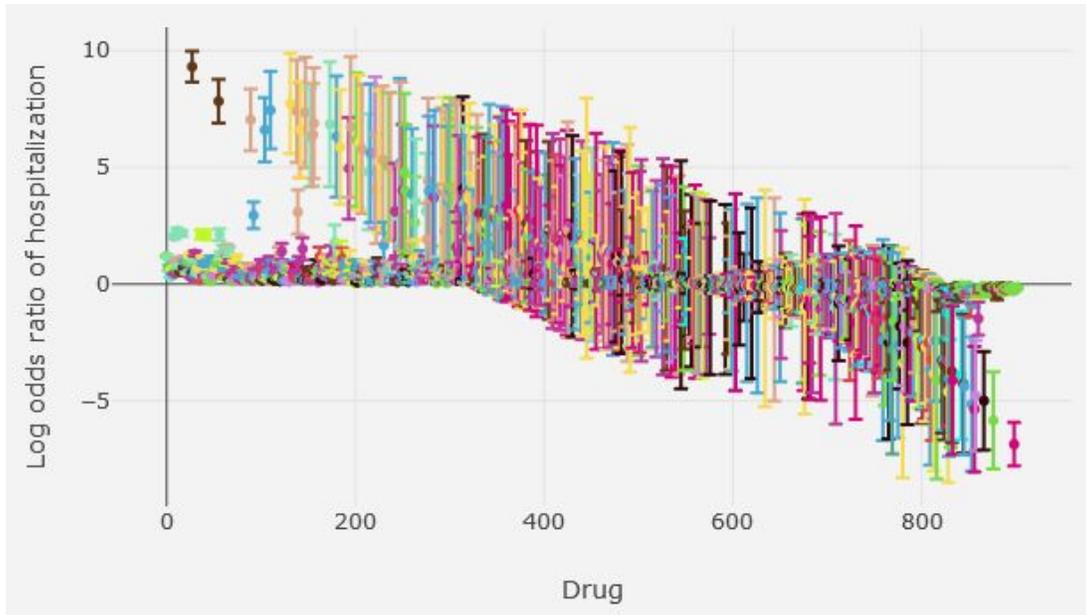

*Figure 5: All bootstrapped logistic regression parameters plus intervals denoting one standard error in each direction, based on 2000 bootstrap resamples. The figure is generated in Supplementary Notebook 4.*

## Resource requirements and execution times

We measured the resource requirements and execution times for training the model on a server with an Intel Xeon 4-core 2-way HT 3.8GHz processor, 64 GB RAM, an NVIDIA GTX 1080Ti with 11GB VRAM, and a 500GB SSD.

Preprocessing a dataset takes about 2 minutes. Converting to Parquet reduces storage footprint for the datasets from 14.5 GB to 3.5 GB. These are therefore much faster to load from disk and have increased read performance due to the indexing.

The training time for logistic regression is 6.1 seconds. Most of our other exploratory statistical analyses have similar short execution times. However, bootstrap with 2000 iterations on a single core takes 15 hours and 12 minutes. All bootstrap resampling procedures can be run in parallel, so we can reduce execution time by using multiple cores. Doing a resampling procedure for error estimates may not be necessary here, but this does provide some suggestion as to what to expect doing something seriously compute intensive at this scale.

## Conclusions and future work

We have designed a system which enables pharmacoepidemiological data exploration with interactive response times. It uses Apache Spark to enable analyses to scale to drug consumption and hospitalization data in 700 000 Norwegian elders, with 60 million prescriptions and 1.9 million hospitalizations in a three-year period. We believe it will scale to data even larger than the sources we have had available.

We demonstrated the usefulness of the systems by exploring the association between drug usage patterns and hospitalizations among Norwegian elders. Using bootstrapping with linear regression, we found a prioritized list of drugs that are over- and underrepresented in the hospitalized group. This list can be used for further analyses, but a large number of drugs had high variance so more data is necessary to decrease uncertainty in these estimates.

The systems can be improved by replacing the spylon-kernel, which is not actively maintained, with Apache Toree (https://toree.apache.org/). We also believe the backend can be moved to a cloud service that provides managed Spark to take advantage of the continuous improvements and elasticity of cloud services.

We believe that our system can be useful in pharmacoepidemiologic research in the future. The preprocessing can be further refined by improving data cleaning procedures. Furthermore, new features such as disease classifications can be added to enable machine learning models to learn more complex patterns between drug consumption and outcomes. Other outcomes such as any adverse drug reactions can be used and larger multi-national data can also be used for such analyses.

The code is open sourced under an MIT license and available at: https://github.com/uit-hdl/norpd_prescription_analyses

# Acknowledgments

The data used in this project belongs to a project with approval from the Regional Committee for Medical and Health Research Ethics (2014/2182) and data has been handled in accordance with licence granted by the Norwegian Data Protection Authority.

The data used from the Norwegian Prescription Database and Norwegian Patient Registry are from a grant funded by the Northern Norway Regional Health Trust.

# References

[1] B. L. Strom, S. E. Kimmel, and S. Hennessy, *Textbook of Pharmacoepidemiology*. John Wiley & Sons, 2013.
[2] M. Pirmohamed *et al.*, "Adverse drug reactions as cause of admission to hospital: prospective analysis of 18 820 patients," *BMJ*, vol. 329, no. 7456, pp. 15–19, Jul. 2004.
[3] K. Svendsen, K. H. Halvorsen, S. Vorren, H. Samdal, and B. Garcia, "Adverse drug reaction reporting: how can drug consumption information add to analyses using spontaneous reports?," *Eur. J. Clin. Pharmacol.*, vol. 74, no. 4, pp. 497–504, Apr. 2018.
[4] C. Pierfitte, B. Bégaud, R. Lagnaoui, and N. D. Moore, "Is reporting rate a good predictor of risks associated with drugs?," *Br. J. Clin. Pharmacol.*, vol. 47, no. 3, pp. 329–331, Mar.

1999.
[5] N. Tavassoli, M. Lapeyre-Mestre, A. Sommet, and J.-L. Montastruc, "Reporting rate of adverse drug reactions to the French pharmacovigilance system with three step 2 analgesic drugs: dextropropoxyphene, tramadol and codeine (in combination with paracetamol)," *Br. J. Clin. Pharmacol.*, vol. 68, no. 3, pp. 422–426, Sep. 2009.
[6] M. Zaharia *et al.*, "Apache Spark: a unified engine for big data processing," *Commun. ACM*, vol. 59, no. 11, pp. 56–65, Oct. 2016.
[7] M. Abadi *et al.*, "TensorFlow: A System for Large-scale Machine Learning," in *Proceedings of the 12th USENIX Conference on Operating Systems Design and Implementation*, Berkeley, CA, USA, 2016, pp. 265–283.
[8] F. Pedregosa *et al.*, "Scikit-learn: Machine Learning in Python," *J Mach Learn Res*, vol. 12, pp. 2825–2830, Nov. 2011.
[9] T. J. Oscanoa, F. Lizaraso, and A. Carvajal, "Hospital admissions due to adverse drug reactions in the elderly. A meta-analysis," *Eur. J. Clin. Pharmacol.*, vol. 73, no. 6, pp. 759–770, Jun. 2017.
[10] B. Efron, "Bootstrap Methods: Another Look at the Jackknife," in *Breakthroughs in Statistics: Methodology and Distribution*, S. Kotz and N. L. Johnson, Eds. New York, NY: Springer, 1992, pp. 569–593.

## Supplementary materials

[**Supplementary Notebook 1**](): Ingest the general population prescription data set from NorPD.

[**Supplementary Notebook 2**](): Compare the distribution of drug consumption between live and dead patients by counting occurrence of drugs. The results is a Panda dataframe with drugs sorted by their estimated overrepresentation.

[**Supplementary Notebook 3**](): Plot the top 5 overrepresented drugs in dead patients using the dataframe from Supplementary Notebook 2.

[**Supplementary Notebook 4**](): Code and interactive visualizations for the bootstrapped logistic regression analysis for finding over- and under-represented drugs in hospitalization.